\documentclass[aps,prl, twocolumn,superscriptaddress,amsfonts,a4paper,showpacs]{revtex4}
\usepackage{amsmath}
\usepackage{bm}
\usepackage{graphicx}
\def\BE {\begin{equation}}
\def\EE {\end{equation}}
\def\BEA {\begin{eqnarray}}
\def\EEA {\end{eqnarray}}
\def\BES {\begin{subequations}}
\def\EES {\end{subequations}}
\def\BA {\begin{array}}
\def\EA {\end{array}}
\def\NN {\nonumber}
\def\NN {\nonumber}
\def\ep {\varepsilon}

{

\begin{document}

\title{Adiabatic quantum search with atoms in a cavity driven by lasers}

\author{D. Daems}
\affiliation{QuIC, Ecole Polytechnique,
Universit\'e Libre de Bruxelles, 1050 Bruxelles, Belgium}
\email{ddaems@ulb.ac.be}

\author{S. Gu\'erin}
\affiliation{Institut Carnot de Bourgogne UMR 5209 CNRS,
Universit\'e de Bourgogne, BP 47870, 21078 Dijon, France}
\email{sguerin@u-bourgogne.fr}

\abstract{We propose an implementation of the quantum search
algorithm of a marked item in an unsorted list of $N$ items
by adiabatic passage in a cavity-laser-atom system. We use an
ensemble of $N$ identical three-level atoms trapped in a
single-mode cavity and driven by two lasers. In each atom, the
same level represents a database entry. One of the atoms is marked
by having an energy gap between its two ground states. Appropriate
time delays between the two laser pulses allow one to populate the
marked state starting from an initial entangled state within a
decoherence-free adiabatic subspace. The time to achieve such a
process is shown to exhibit the Grover speedup $\sqrt{N}$. } }

 \maketitle One typical problem
of quantum computation concerns the search of a marked entry in an
unsorted database by accessing it a minimum number of times. The
Grover algorithm \cite{grover} achieves this task quadratically
faster than any classical algorithm. It is formulated in terms of
a series of quantum gates applied to a quantum register consisting
of a collection of qubits encoding the database entries. An
initial uniform superposition $|w\rangle$, independent of the
searched state $|m\rangle$, is rotated step by step  under the
action of appropriate gates. The searched state is exhibited by an
oracle function which checks if a proposed input is the searched
state, returning for instance 1 in this case and 0 otherwise. The
number of steps grows as $N^{1/2}$ with $N$ the database size,
whereas a classical algorithm requires on average $N/2$ calls.
This quantum circuit algorithm has been tested experimentally for
two qubits ($N$=4) by several techniques resting on NMR
\cite{nmr1,nmr2}, optics \cite{optics1,optics2} and trapped ions
\cite{ions}. There have also been proposals of experimental
implementations using cavity QED where the quantum gate dynamics
is provided by a cavity-assisted collision \cite{qed1} or by
a strong resonant classical field \cite{qed2}.

A time continuous version of the Grover algorithm has been
proposed by Fahri and Gutmann  \cite{farhi} who, instead of using
an explicit oracle, mark the searched state with an energy $E$
while the others  are degenerate with energy 0, and use a driving
Hamiltonian that leads continuously the initial state to the
marked one. Choosing an Hamiltonian $V=E|w\rangle\langle w|$ to
drive the free system $H_0=E|m\rangle\langle m|$, they have shown
that a
 Rabi-like half-cycle leads to the target marked state in a time
growing as $N^{1/2}/E$. Note that the Grover speedup is quadratic,
independently of any increase of $E$ with $N$ which would simply
amount to renormalizing the time. An experimental realization of
this analog Grover algorithm has been performed by NMR \cite{nmr3}
in a setting where a quadrupolar coupling  makes a spin 3/2
nucleus a two-qubit system ($N$=4).

Adiabatic versions of the time continuous Grover algorithm have
been proposed \cite{adiabatic1,adiabatic2,roland} mainly to take
advantage of the robustness of adiabatic passage with respect to
fluctuations of the external control fields as well as to the
imperfect knowledge of the model.
%in order to achieve a complete transfer of
%population which, in contrast to the Rabi oscilations of the
%preceding case, is robust and can be maintained in time.
They have been formulated with a Hamiltonian which connects
adiabatically the initial ground superposition $|w\rangle$ to the
marked state $|m\rangle$ through an avoided crossing:
$H=[1-u(t)]H_i+u(t)H_f$ where $H_i=I-|w\rangle\langle w|$,
$H_f=I-|m\rangle\langle m|$, and $u(t)$ is a function of time growing
from 0 to 1. Roland and Cerf \cite{roland} have shown that only a
specific speed of the dynamics controlled by $u(t)$ allows one to
achieve the transfer to the marked state in a time growing as
$N^{1/2}$.

\begin{center}
\begin{figure}
\includegraphics[scale=0.55]{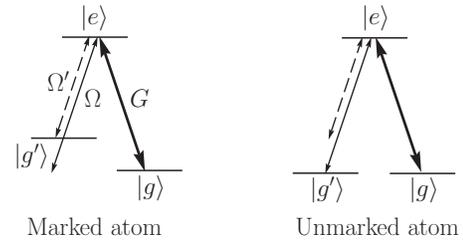}
\caption{Linkage pattern for the individual atoms. The unmarked
atoms have two degenerate ground states $|g\rangle$ and
$|g'\rangle$. One atom is marked with the state $|g'\rangle$
shifted. The laser of Rabi frequency $\Omega'$ (resp. $\Omega$) is
resonant with the $g'-e$ transition for the marked (resp.
unmarked) atom(s). The cavity of Rabi frequency $G$ is resonant
with the $g-e$ transition.} \label{atoms}
\end{figure}
\end{center}

\begin{center}
\begin{figure}
\includegraphics[scale=0.55]{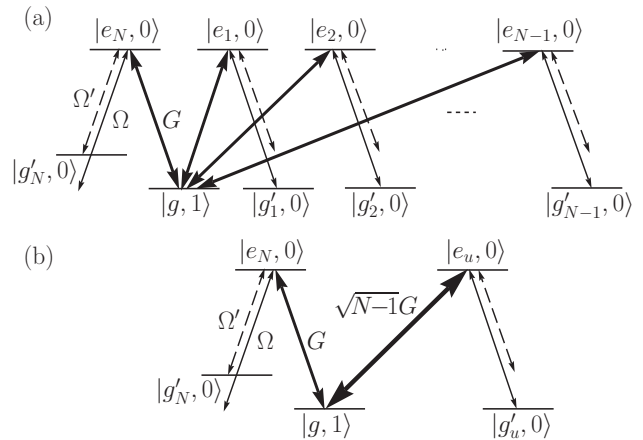}
\caption{(a) Coupling scheme in the cavity. The cavity $G$, laser
$\Omega$, laser $\Omega'$ Rabi frequency are featured by
respectively thick, thin, and dashed arrows. (b) Equivalent scheme
where the states $|g'_i,0\rangle$ (resp. $|e_i,0\rangle$),
$i=1,N-1$, of frame (a) form the collective unmarked ground state
$|g'_u,0\rangle$ (resp. excited state $|e_u,0\rangle$). The
effective cavity Rabi frequency to the collective unmarked excited
state is $\sqrt{N-1}G$.} \label{FigStar}
\end{figure}
\end{center}

In this paper, we show an implementation of the adiabatic Grover
algorithm based on a physical system, which is in principle
scalable. This is, to our knowledge, the first proposed
implementation of the adiabatic Grover algorithm. It is formulated
with an Hamiltonian $H=H_0+V(t)$ where $H_0$ is considered as an
oracle, and given, while  $V(t)$ is slowly varying in time and
such that there is an instantaneous eigenvector adiabatically
connecting the initial superposition $|w\rangle$ to the marked
state $|m\rangle$. We use an ensemble of $N$ identical three-level
atoms trapped in a single-mode cavity of coupling frequency $G$,
and driven by two lasers of Rabi frequencies $\Omega$ and
$\Omega'$.
The atomic levels are in a $\Lambda$ configuration with two ground
states $|g\rangle$ and $|g'\rangle$  coupled to the excited state
$|e\rangle$ by, respectively, the cavity and the two lasers (see
Fig. \ref{atoms}). This can be realized in practice by considering
 Zeeman states and  laser and cavity fields of appropriate polarizations.
The states $|g'\rangle$ of the $N$ atoms are considered as the
database entries. The energy of the state $|g'\rangle$ of the
marked atom is shifted by an amount $\delta$ with respect to that
of the unmarked atoms, which is set to zero.
The states $|g\rangle$ allow for the coupling of all the atoms
through the exchange of a single photon with the cavity (see Fig.
\ref{FigStar}a).
The initial state we shall start from is the entangled state
$|w\rangle\equiv
|g^\prime,0\rangle\equiv(1/\sqrt{N})\sum_{j=1}^N|g'_j,0\rangle$
featuring a collective single-photon atomic excitation. Note that
the label of each atom is added as a subscript $1,\cdots,N$ and
chosen so that the marked atom has tag $N$. Such a state can be
prepared for instance before the marking of the atom using the
stimulated Raman adiabatic passage (STIRAP) technique
\cite{STIRAP}, exactly as shown in \cite{Fleisch} to store
single-photon quantum states.

The process we introduce here allows one to drive
 adiabatically the population of the entangled state $|g^\prime,0\rangle$,
 which corresponds to the superposition of
both the marked state $|m\rangle\equiv |g'_N,0\rangle$ and an unmarked collective
state $|g'_u,0\rangle$ that we introduce below, to the single marked state
$|g^\prime_N,0\rangle$ (see Fig. \ref{FigStar}b).
%reminiscent of the above STIRAP technique. It can be referred to
This process will be referred to as an \textit{inverse fractional
stimulated Raman adiabatic passage} (if-STIRAP) since it is a time
inversion of the so-called fractional STIRAP (f-STIRAP) which transfers
the population from a single state to a superposition of state
\cite{fstirap}. This is implemented by first switching on $\Omega$
and $\Omega'$ together and next switching off $\Omega'$ before
$\Omega$. The time to achieve such a process will then be shown to
grow as $\sqrt{N}$ in order to satisfy adiabaticity.

 The Hamiltonian describing the system of $N$ atoms is
\BE
 H_0=\delta |g_N\rangle  \langle g_N| + \omega \sum_{j=1}^N |e_j\rangle \langle e_j| .
\EE
 We consider a cavity mode of frequency $\omega$ and coupling strength $G$ together with
 two lasers of frequencies $\omega$, $\omega-\delta$
 and  pulse shapes $\Omega(t)$, $\Omega^\prime(t)$ which do not grow with $N$.
 The resonant driving provided by the atoms-cavity-laser system is described by
\BEA
 V&=&\omega a^\dagger a  + G a \sum_{j=1}^N |e_j\rangle  \langle g| \NN\\
 &+&  \left[\Omega e^{i \omega t} +\Omega^\prime e^{i (\omega-\delta) t}\right]
 \sum_{j=1}^{N}  |g^\prime_j \rangle \langle e_j| +\rm{h. c.}  \label{HD}
\EEA
The full Hamiltonian $H=H_0+V$ has a photonic block diagonal structure.
Each block is labeled by the number $k$ of photons in the cavity when the $N$ atoms are in
their ground state $|g\rangle$. The corresponding multipartite
state $|g_1 \cdots  g_N  \rangle \otimes |k \rangle$ is denoted
$|g,k\rangle$. All the states which are connected to $|g,k\rangle$
span a subspace whose projection operator is $P_k$. As  each block
is decoupled under $H$ from the other ones, $H=\sum_{k=0} P_k H
P_k$, we shall focus on the block $P_1 H
P_1$ associated with a single photon
in the cavity and show that it allows us to implement an adiabatic
Grover search algorithm.

The  multipartite state  $|g,1\rangle$ is connected by (\ref{HD})
to exactly two families of states as illustrated in Fig. \ref{FigStar}a.
Upon absorption of the cavity photon, the excited state
$|e\rangle$ of any of the $N$ atoms, say atom $j$, can be reached
while the other atoms remain in their ground state $|g\rangle$;
the corresponding multipartite state  is $|e_j,0\rangle \equiv
|g_1 \cdots  g_{j-1} e_{j} g_{j+1}   \cdots \rangle \otimes
|0\rangle$. The state $|e_j,0\rangle$ can also be reached by
absorption of one laser photon when any of the atoms, say atom
$j$, is in the ground state $|g^\prime\rangle$ while the other
atoms are in the ground state $|g\rangle$:
$|g^\prime_j,0\rangle\equiv |g_1 \cdots  g_{j-1} g^\prime_{j}
g_{j+1}   \cdots g_N \rangle \otimes  |0\rangle$.
%The states $\left\{|g,1\rangle,|g^\prime_k,0\rangle, |e_k,0\rangle\right\}_{k=1}^N$

In order to remove the oscillatory time dependence introduced by
the laser we consider atomic states which are dressed by laser and
cavity photons, and use the resonant transformation $R=e^{-i
\delta t} |g^\prime_N,0\rangle \langle g^\prime_N,0|+e^{-i \omega
t} |g,1\rangle \langle g,1|+
\sum_{j=1}^{N}\left(|g^\prime_j,0\rangle \langle
g^\prime_j,0|+e^{-i \omega t} |e_j,0\rangle \langle
e_j,0|\right)$.

As we shall see, the states which are relevant for the Grover
search are the $N-1$ states $|g^\prime_j,0\rangle$, which are
unmarked, and the state $|g^\prime_N,0\rangle$ which is marked.
Among the unmarked atoms, none should play a privileged role.
Hence we shall consider them collectively and label the
corresponding state with a subscript $u$. We  rewrite this
Hamiltonian in a new basis which features the uniform
superposition of the unmarked ground states \BE
|g^\prime_u,0\rangle=\frac{1}{\sqrt{N-1}}\sum_{j=1}^{N-1}|g^\prime_j,0\rangle,
\EE
 and the uniform superposition of the excited states associated
with the unmarked atoms
\BE
|e_u,0\rangle=\frac{1}{\sqrt{N-1}}\sum_{j=1}^{N-1}|e_j,0\rangle.
\EE
This is achieved with the unitary transformation
$W=|g^\prime_N,0\rangle \langle g^\prime_N,0|+ |e_N,0\rangle
\langle e_N,0|+ |g,1\rangle  \langle g,1|+ \sum_{l,j=1}^{N-1}
U_{l,j}\left(|g^\prime_l,0\rangle  \langle
g^\prime_j,0|+|e_l,0\rangle  \langle e_j,0|\right)$ where $U$ is
any unitary matrix such that the $(N-1)$ elements of one of its
columns are equal. In this new basis, the part of the Hamiltonian
restricted to the subspace  spanned by the states
$|g^\prime_u,0\rangle,
|g^\prime_N,0\rangle,|g,1\rangle,|e_u,0\rangle,|e_N,0\rangle$ is
decoupled from the rest.
It reads
\BEA
H_1=\left(
\BA{cc}
0 & V_{\ell}\\
V_{\ell}^\dagger & V_{\rm c} \EA \right) ,
\EEA 
with 
\BEA
V_{\ell}=\left( \BA{ccc}
0&\Sigma& 0\\
0&0 & \Sigma^\prime \EA \right), \; V_{\rm c}=\left( \BA{ccc}
0 &  \sqrt{N-1} G &  G\\
 \sqrt{N-1} G & 0& 0 \\
G & 0 & 0 \EA \right) .\NN 
\EEA 
and $\Sigma=\Omega+e^{-i \delta t}
\Omega^\prime$, $\Sigma^\prime=\Omega^\prime+e^{i \delta t}
\Omega$. This Hamiltonian, whose derivation is exact, is represented in Fig. \ref{FigStar}b.

By means of a unitary transformation $T$, we diagonalize the block
$V_{\rm c}$ which admits the
 eigenvalues  $\gamma_0=0$ and $\gamma_\pm =\pm \sqrt{N} G $ whose associated eigenvectors read
\BEA
|\gamma_0\rangle&= &\sqrt{1-\frac{1}{N}} |e_N,0 \rangle -\frac{1}{\sqrt{N}}|e_u,0 \rangle\\
|\gamma_\pm\rangle&=&\frac{1}{\sqrt{2}}\left( \frac{1}{\sqrt{N}} |e_N,0 \rangle +\sqrt{1-\frac{1}{N}}|e_u,0 \rangle  \pm  |g,1 \rangle\right).\NN
\EEA

The new Hamiltonian reads, in the basis $|g^\prime_u,0\rangle,|g^\prime_N,0\rangle,|\gamma_0\rangle,|\gamma_+\rangle,|\gamma_-\rangle$ :
\BEA
T^\dagger H_1 T=\left(
\BA{cc}
A & B\\
  B^\dagger & C
\EA
\right),
\EEA
with
\BEA
A&=&\frac{1}{\sqrt{N}}\left(
\BA{ccc}
0 \ & 0  \  &  -\Sigma\\
0 \ & 0  \  & \sqrt{N-1}\Sigma^\prime\\
 -\Sigma^*  & \sqrt{N-1}\, {\Sigma^\prime}^* & 0
\EA
\right)\NN\\
B&=&\frac{1}{\sqrt{2N}}\left(
\BA{ccc}
\sqrt{N-1}\Sigma & \sqrt{N-1}\Sigma\\
\Sigma^\prime &  \Sigma^\prime\\
0&0
\EA
\right) \NN\\
C &=&\sqrt{N}\left(
\BA{cc}
 G&  0\\
0& - G
\EA
\right).
\EEA

The time evolution of the non-resonant components in $\Sigma$ and
$\Sigma^\prime$ is much faster than the evolution of $\Omega$ and
$\Omega^\prime$ which occurs over a time scale $\mathcal{T} \gg
\delta^{-1}$. Hence it is justified to replace these contributions
by their vanishing average values over times $ \delta^{-1} \ll
\tau \ll \mathcal{T}$: $\overline{\Sigma} \simeq \Omega$ and
$\overline{\Sigma^\prime}\simeq \Omega^\prime$, where
$\overline{f}(t)=\frac{1}{\tau} \int_t^{t+\tau}du f(u)$. This is
the resonant approximation. Similarily, the unitary evolution of
$C$ is much faster than that of $A$ if $\Omega_{\rm peak}/ N G \ll 1$ since
the respective eigenvalues scale as $\sqrt{N} G$  and
$\Omega_{\rm peak}/\sqrt{N}$ where $\Omega_{\rm peak}$ is the peak amplitude of the
pulse $\Omega$. Upon performing an adiabatic elimination we thus
obtain an effective Hamiltonian $H_{{\rm
eff}}=\overline{A}-\overline{B} C^{-1} \overline{B}^\dagger$ which
contains all the contributions up to order $(\Omega_{\rm peak}/ N G)^4$. In
the basis $|g^\prime_N,0\rangle, |\gamma_0\rangle,
|g^\prime_u,0\rangle$, the effective Hamiltonian
reads \BEA H_{{\rm eff}}=\frac{1}{\sqrt{N}}\left( \BA{ccc}
0   &  \sqrt{N-1} \Omega^\prime  & 0 \\
\sqrt{N-1}  \Omega^\prime & 0  \  &  - \Omega \\
 0 & - \Omega  &  0 \\
\EA
\right). \label{Heff}
\EEA

Our aim is to transfer adiabatically the population from an
initial state $|g^\prime,0\rangle$ which gives no privileged role to any of the $N$
states $|g^\prime_j,0\rangle$ to a final state  which coincides
with the marked state $|g^\prime_N,0\rangle$ in a time which
scales as $\sqrt{N}$. The population transfer mechanism is most
easily revealed in the basis of the instantaneous eigenstates  of
$H_{{\rm eff}}(t)$ 
\BEA
|0\rangle(t) &=& \cos \theta(t) \ |g^\prime_N,0\rangle
-\sin \theta(t) \ |g^\prime_u,0\rangle\label{adiab}\\
|\pm\rangle(t) &=& \frac{1}{\sqrt{2}}\left(\sin \theta(t) \
|g^\prime_N,0\rangle +\cos \theta(t) \ |g^\prime_u,0\rangle \pm
|\gamma_0 \rangle \right),\NN 
\EEA 
pertaining to the eigenvalues 0
and $\pm \Lambda(t)$ where 
\BEA
\Lambda(t)=\frac{1}{\sqrt{N}}\sqrt{(N-1){\Omega^\prime}^2(t)+\Omega^2(t)}.
\label{Lambda} 
\EEA 
The instantaneous angle $\theta(t)$ is defined
through the relation 
\BEA
\tan \theta(t) &=& -\sqrt{N-1}
\frac{\Omega^\prime(t)}{\Omega(t)}. 
\label{tan} 
\EEA 
Requiring the
instantaneous eigenstate (\ref{adiab}) to coincide at the initial
time $t_i$ with  the uniform superposition
\BE
|g^\prime,0\rangle=\frac{1}{\sqrt{N}}  |g^\prime_N,0 \rangle +
\sqrt{1-\frac{1}{N}}  |g^\prime_u,0 \rangle,
\EE
and at the final
time with the marked state $|g^\prime_N,0\rangle$ entails that
\BE
\label{tanif}
\tan \theta(t_{\rm i})= -\sqrt{N-1},\quad \tan
\theta(t_{\rm f})= 0 .
\EE
This implies that the two pulses must
be switched on simultaneously, $\Omega^\prime(t_{\rm
i})=\Omega(t_{\rm i})$ and that the pulse $\Omega^\prime$ is to be
turned off before $\Omega$.
In the adiabatic representation (\ref{adiab}), the Hamiltonian (\ref{Heff}) reads
\BEA
H_{{\rm eff}}^{\rm ad}= \left(
\BA{ccc}
\Lambda  & \frac{i}{\sqrt{2}} \dot \theta  & 0 \\
-\frac{i}{\sqrt{2}} \dot \theta  & 0  & -\frac{i}{\sqrt{2}} \dot \theta \\
 0 & \frac{i}{\sqrt{2}} \dot \theta &  -\Lambda \\
\EA \right), 
\EEA 
where  $\dot \theta =\frac{1}{1+\tan^2 \theta}
\frac{d}{dt} \tan \theta$. In the adiabatic regime, the
transitions between instantaneous eigenstates are negligible. This
will be achieved if the Hamiltonian varies sufficiently slowly in
time so as to keep $\dot \theta \ll \Lambda$. On the other hand,
we wish to control the process duration and, in particular, to
prevent it from becoming arbitrary large. For that purpose, as
proposed in \cite{roland}, we choose to require $\dot \theta$ and
$\Lambda$ to be in a constant (small) ratio $\ep$ at all times,
independently of $N$: 
\BE 
\label{epLambda} \dot \theta = \ep
\Lambda. 
\EE 
This choice is adopted for the sake of clarity, since
in this case the scaling can be analytically determined and proven
to scale as $\sqrt{N}$. We have investigated the  robustness of
this approach.

Given a laser pulse $\Omega$, this equation will allow us to
determine the pulse $\Omega^\prime$ which is needed to remain in
the instantaneous eigenstate $|0\rangle(t)$ with a probability
larger than $1 -\ep^2$ throughout the process, starting from the
uniform superposition $|0\rangle(t_{{\rm i}})=|w\rangle$,  and
ending up in the marked state $|0\rangle(t_{{\rm
f}})=|g^\prime_N,0\rangle$ after some time
$\mathcal{T}=t_{{\rm f}}-t_{{\rm i}}$. Indeed, rewriting
(\ref{Lambda}) with (\ref{tan}) as
$\Lambda=\frac{1}{\sqrt{N}}\sqrt{1+\tan^2 \theta}  \Omega$, we
obtain from (\ref{epLambda}) a differential equation for $\tan
\theta$, {\em i. e.}, for the ratio $\Omega^\prime/\Omega$. Its
solution satisfying the initial condition (\ref{tanif}) reads
\BE
\frac{\Omega^\prime(t)}{\Omega(t)}=\frac{1-\frac{\ep {\mathcal
A}(t)}{\sqrt{N-1}}}{\sqrt{1+\ep {\mathcal A}(t)\{2\sqrt{N-1}-\ep
{\mathcal A}(t)\}}} ,
\label{rho}
\EE
where we define ${\mathcal
A}(t)\equiv\int_{t_{{\rm i}}}^t du \Omega(u)$.
The process duration $\mathcal{T}$ is obtained implicitly upon specifying that
at time $t_{{\rm f}}$ the ratio on the left side of (\ref{rho})
vanishes: 
\BE 
\ep   {\mathcal A}(\mathcal{T})=\sqrt{N-1}. 
\EE
Expressing the total area of the pulse $\Omega$ in terms of its
average amplitude $\overline {\Omega}$, ${\mathcal
A}(\mathcal{T})=\overline {\Omega} \mathcal{T}$, we arrive at
%the process duration
\BE  
\overline {\Omega}\mathcal{T}=\frac{\sqrt{N-1}}{\ep}. 
\EE
This shows that, for a constant average amplitude (i.e.
independent of $N$), the duration scales as $\sqrt{N}$. Note that
we can equivalently increase the average amplitude as $\sqrt{N}$
for a constant time $\mathcal{T}$. We can determine from
(\ref{rho}) that $\overline {\Omega}^{\prime}\mathcal{T}$ grows as
$(\sqrt{N}-1)/\sqrt{N-1}$, {\em i.e.}, as $O(N^0)$.

\begin{center}
\begin{figure}
\includegraphics[scale=0.55]{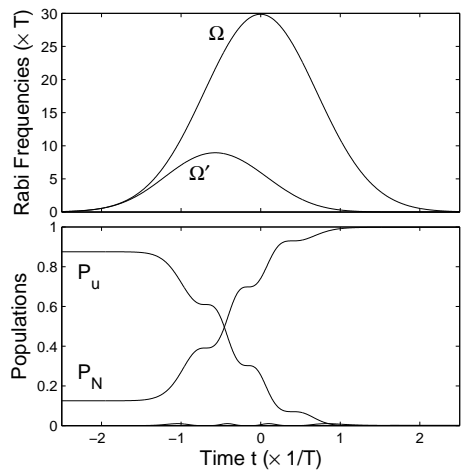}
\caption{Numerical dynamics with the effective Hamiltonian
(\ref{Heff}) for $N=8$, $\varepsilon=0.05$ and a Gaussian Rabi
frequency $\Omega(t)=\Omega_{\rm peak}e^{-(t/T)^2}$ with
$\Omega_{\rm peak}T=\sqrt{N-1}/(\varepsilon\sqrt{\pi})$ [in order
to satisfy (\ref{tanif}) and (\ref{epLambda})].
$\Omega^{\prime}(t)$ is determined from (\ref{rho}). Top: Rabi
frequencies. Bottom: Populations of the collective unmarked state
$P_u(t)\equiv|\langle g^\prime_u,0|\phi\rangle(t)|^2$, and of the
marked state $P_N(t)\equiv|\langle
g^\prime_N,0|\phi\rangle(t)|^2$,
where $|\phi\rangle(t)$ is the dynamical state vector.}
\label{Dynamics}
\end{figure}
\end{center}

Figure \ref{Dynamics} displays the pulses and the population dynamics resulting from 
(\ref{rho}) with $N=8$, 
$\varepsilon=0.05$  and a Gaussian pulse $\Omega$ of
characteristic duration $T$. 
As predicted, the transfer to the
marked state is very efficient. 
The very low transient population
in the excited states stems from the fact that the dynamics is
expected to remain in the instantaneous decoherence-free
eigenstate $|0\rangle(t)$ in the adiabatic limit. Notice that the
choice (\ref{epLambda}), which leads to the seemingly complicated
pulse relation (\ref{rho}), gives in practice a simple smooth
bell-shaped pulse (see Fig. \ref{Dynamics}). We have numerically
checked that the efficiency of the transfer is, as expected,
preserved for a wide range of $N$ with $\Omega_{\rm peak} T$
growing as $\sqrt{N}$ and an almost constant $\Omega^{\prime}_{\rm
peak} T$.

In conclusion, we have proposed the first physical implementation
of the adiabatic Grover search using a cavity-laser-atom system
and robust processes related to STIRAP. The calculation has been
conducted with pulses based on the constraint (\ref{epLambda})
that has allowed us to prove the scaling analytically. We have
checked the robustness of the $\sqrt{N}$ scaling by numerical
simulations using other less restrictive adiabatic pulse shapes
satisfying (\ref{tanif}).

The authors are grateful to N. J. Cerf and H.-R. Jauslin for
useful discussions and acknowledge the support from the EU
projects COVAQIAL and QAP, from the Belgian government programme
IUAP under grant V-18, and from the Conseil R\'egional de
Bourgogne.


\begin{thebibliography}{99}
%\vspace{-2mm}
\bibitem{grover} L. K. Grover, Phys. Rev. Lett, {\bf 79}, 325 (1997).
\bibitem{nmr1} I. L. Chuang, N. Gershenfeld, and M. Kubinec, Phys. Rev. Lett. {\bf 80}, 3408 (1998).
\bibitem{nmr2} J. A. Jones, M. Mosca, and R. H. Hansen, Nature (London) {\bf 393}, 344 (1998).
\bibitem{optics1} N. Bhattacharya, H. B. van Linden van den Heuvell, and R. J. C. Spreeuw, Phys. Rev. Lett. {\bf 88}, 137901 (2002).
\bibitem{optics2} P. Walther, K. J. Resch, T. Rudolph, E. Schenck, H. Weinfurter, V. Vedral, M. Aspelmeyer, and A. Zeilinger, Nature (London) {\bf 434}, 169 (2005).
\bibitem{ions} K.-A. Brickman, P. C. Haljan, P. J. Lee, M. Acton, L. Deslauriers, and C. Monroe, Phys. Rev. A {\bf 72}, 050306(R) (2005).
\bibitem{qed1} F. Yamaguchi, P. Milman, M. Brune, J. M. Raimond, and S. Haroche, Phys. Rev. A {\bf 66}, 010302(R) (2002).
\bibitem{qed2} Z. J. Deng, M. Feng, and K. L. Gao, Phys. Rev. A {\bf 72}, 034306 (2005).
\bibitem{farhi} E. Farhi and S. Gutmann, Phys. Rev. A \textbf{57}, 2403 (1998).
\bibitem{nmr3} V. L. Ermakov and B. M. Fung, Phys. Rev. A {\bf 66}, 042310 (2002).
\bibitem{adiabatic1} E. Farhi, J. Goldstone, S. Gutmann, and M. Sipser,
e-print quant-ph/0001106.
\bibitem{adiabatic2} A. M. Childs, E. Farhi, and J. Preskill,  Phys. Rev. A {\bf 65}, 012322 (2001).
\bibitem{roland} J. Roland and N. J. Cerf, Phys. Rev. A \textbf{65}, 042308 (2002).
\bibitem{STIRAP} N. V. Vitanov, T. Halfmann, B. W. Shore, and K. Bergmann,
Annu. Rev. Phys. Chem. \textbf{52}, 763 (2001).
\bibitem{Fleisch} M. Fleischhauer and M. D. Lukin, Phys. Rev. A \textbf{65}, 022314 (2002).
\bibitem{fstirap} N. V. Vitanov, K.-A. Suominen, and B. W. Shore, J. Phys. B {\bf 32}, 4535 (1999).
\end{thebibliography}
\end{document}